\documentclass[12pt]{iopart}
\usepackage{iopams,cite}
\usepackage{graphicx}
\usepackage{amsfonts}
\usepackage{mathrsfs}
\usepackage{epstopdf}
\usepackage{bm}

\newcommand{\epl}{{\it Europhys.\ Lett.\ }}

\newcommand{\prl}{{\it Phys.\ Rev.\ Lett.\ }}
\newcommand{\pra}{{\it Phys.\ Rev.\ A\ }}
\newcommand{\prb}{{\it Phys.\ Rev.\ B\ }}
\newcommand{\pre}{{\it Phys.\ Rev.\ E\ }}
\newcommand{\rmp}{{\it Rev.\ Mod.\ Phys.\ }}
\newcommand{\pr}{{\it Phys.\ Rep.\ }}

\newcommand{\pla}{{\it Physics Lett. A\ }}

\newcommand{\ajp}{{\it Am.\ J.\ Phys.\ }}
\newcommand{\njp}{{\it New\ J.\ Phys.\ }}

\begin{document} 

\title{Scale-free relaxation of a wave packet in a quantum well with power-law tails}

\author{Salvatore Miccich\`e$^1$, Andreas Buchleitner$^2$, Fabrizio Lillo$^{1,3,4}$, Rosario N. Mantegna$^1$, 
Tobias Paul\footnote{Now at AREVA NP GmbH, Germany}$^{5}$ and Sandro Wimberger$^5$}

\address{$^1$Dipartimento di Fisica, Universit\`a degli Studi di Palermo, Viale delle Science Ed. 18, 90128 Palermo, Italy}
\address{$^2$Physikalisches Institut, Albert-Ludwigs-Universit\"at Freiburg, Hermann-Herder-Str. 3, 79104 Freiburg, Germany}
\address{$^3$Santa Fe Institute, 1399 Hyde Park Road, Santa Fe, NM 87501, USA}
\address{$^4$Scuola Normale Superiore di Pisa, Piazza dei Cavalieri 7, 56126 Pisa, Italy}
\address{$^5$Institut f\"ur Theoretische Physik and Center for Quantum Dynamics, Universit\"at Heidelberg, Philosophenweg 19, 69120 Heidelberg, Germany}

\begin{abstract}
We propose a setup for which a power-law decay is predicted to be observable for generic and realistic conditions. The system we study is very simple: 
A quantum wave packet initially prepared in a potential well with (i) tails asymptotically decaying like $\sim x^{-2}$ and (ii) an eigenvalues spectrum 
that shows a continuous part attached to the ground or equilibrium state. We analytically derive the asymptotic decay law from the spectral properties for generic, 
confined initial states. Our findings are supported by realistic numerical simulations for state-of-the-art expansion experiments with cold atoms.
\end{abstract}


\maketitle

\section{Introduction}  \label{intro}

The temporal evolution of an initially localized, quantum mechanical wave packet is of fundamental importance for our 
understanding of the classical and quantum transport, and goes back to Schr\"odinger's early 
attempts to construct wave groups which behave alike classical particles \cite{schroedinger26}. With increasing complexity 
of the potential landscape, which can be due to its topography as well its topology, a panoply of surprising transport phenomena
emerge, from ballistic over sub-, super- or just diffusive \cite{bal,transp} to various types of localized transport \cite{anderson,casati79,cbs,disorder}. 
These phenomena have their respective effective descriptions, e.g. in semiclassical \cite{haake_book}, mesoscopic \cite{imry} or statistical 
\cite{hornberger_vacchini} terms, and manifest in distinct experimental settings, from light-matter interaction \cite{arndt91} over 
fundamental (quantum) optics \cite{light,longhi} to ultracold matter \cite{bdz,billy}, quantum walks 
\cite{qrw} and biochemistry \cite{bio}. On the fundamental level, however, all transport properties are hardwired in the spectral properties of the 
underlying quantum system, and control of the latter implies control of the former. 
Given the stunning experimental control over potential landscapes in variable dimensions, as achieved e.g. in cold matter science over the 
last decade \cite{billy,davidson,toulouse,saw,selim}, we can contemplate to explore the above diversity of quantum transport phenomena for optimal control, 
by tuning the decisive spectral properties. 

In this paper we study the particularly simple, though paradigmatic case of a tunneling escape from a one dimensional potential well. 
The proper choice of the confining potential allows one to induce algebraic rather than exponential decay, for generic, confined initial states. 
As described in references \cite{zoller96,muga}, the crucial ingredient will be an asymptotically scale 
invariant potential decaying to zero as $\sim x^{-2}$, which occurs naturally in systems with dipolar interactions \cite{dipole} or 
anomalous molecular binding potentials \cite{efimov}, and is often associated with quite counterintuitive effects 
(see, e.g., \cite{zoller96,lm00,lmm02,eg06,Miyam04}). 
We thus provide in the next section \ref{theory} a particularly transparent example of a scale-free relaxation process, where the 
equilibrium state is reached only at asymptotically long times \cite{transp,fine}.
The here considered scenario is shown to be robust against unavoidable experimental modifications of the idealized theoretical scenario we 
depart from. In section \ref{numeric} will argue that algebraic, rather than exponential decay, should be observable in modern experiments with expanding Bose-Einstein 
condensates \cite{billy,saw} in engineered potentials \cite{davidson,toulouse,selim}. The final section \ref{conclusions} concludes the paper.

\section{Power-law relaxation} \label{theory}

We start our analysis by considering the two parameter family of 1D, single particle potentials \cite{lmm02}
\begin{eqnarray} \label{eq:1}
                              &&V_S(x)= \left\{
                                                            \begin{array}{r@{\quad , \quad}l}
                                                           - V_0                  & |x| \leq L\\ 
                                                           \frac{V_1}{x^2} & |x| >     L \;.
                                                    \end{array} \right. \label{eq:1a}
\end{eqnarray}
We use dimensionless units in the sequel, measuring actions in units of $\hbar$, and giving the particle unit mass. 
The exact analytical solution of the Schr\"odinger equation for the potential $V_S$ is available \cite{farago,lmm02}. We will however see that, given its asymptotic scaling, the precise form of the confining well is not crucial for the predicted power-law decay of the survival probability. The potential of equation~(\ref{eq:1}) has the same asymptotic behavior as in references \cite{zoller96,muga}, however, differently from the one considered in \cite{muga} it extends over the entire real axis, implying the existence of an equilibrium or ground state.

\subsection{Theoretical predictions}  \label{theory-A}

Equation~(\ref{eq:1a}) ensures that the ground state eigenfunction $\psi_0$ associated with the eigenvalue $E=0$ is continuous with a continuous first derivative. For $|x|>L$ one finds $\psi_0=A_0/x^{\alpha \over{2}}$, with $\alpha=\sqrt{1 + 8 V_1}-1$ \cite{zoller96,lmm02}, what ensures the square integrability for $\alpha>1$. For $|x| \leqslant L$, the eigenfunction of the ground state is $\psi_0=B_0 \,\cos(\sqrt{V_0}\,x)$.  The constants $A_0$ and $B_0$ are set by imposing that $\psi_0$ is normalized to unity and continuous in $x=\pm L$. The continuity of the first derivative in $x=\pm L$ implies the following relation between the three original parameters of the potential
\begin{eqnarray}
                              V_1=L\, \sqrt{V_0}\,\tan\bigl( \sqrt{V_0} L\bigl) \Bigl(1+L \, \sqrt{V_0} \,\tan\bigl( \sqrt{V_0} L\bigl)\Bigl)/2 \,.   
\label{eq:1b}
\end{eqnarray}
Besides the discrete eigenvalue $E=0$, the spectrum of $V_S$ has a continuous component in the range $E>0$, which is attached to the null eigenvalue. Since the potential is even, we can have odd and even solutions to the Schr\"odinger equation. For $|x| > L$ the eigenfunction $\psi_E^{(odd)}$ is a linear combination of Bessel functions $\psi_E^{(odd)}=a_E\,\sqrt{x} J_\nu(\sqrt{E}\,x)+b_E\,\sqrt{x}\, Y_\nu(\sqrt{E}\,x)$ where $\nu=(\alpha+1)/2$. For $|x| \leqslant L$ we find $\psi_E^{(odd)}=d_E\,\sin(\sqrt{V_0+E}\,x)$. The coefficients $a_E$, $b_E$ and $d_E$ are fixed by imposing that $\psi_E^{(odd)}$ and its first derivative are continuous in $x=L$ and that $\psi_E^{(odd)}$ are orthonormalized with a $\delta$-function of the energy: $\int dx \, \psi_E(x) \psi_{E'}(x) = \delta(E-E')$. Similar conditions apply to the even solutions. Further details can be found in \ref{sol}.

Let us now consider the particle prepared in an initial state $\Psi_0(x)=\Psi(x,0)$, given by a linear 
combination of eigenfunctions $\{\psi_0(x), \psi_E(x)\}$, with real 
coefficients $\{a_0, a_E\}$. 
Its time evolution will be given by 
\begin{eqnarray}
                \Psi(x,t)=a_0\,\psi_0(x) + 
                        \int_0^\infty d E\,a_E\,\psi_E(x)\,e^{- i E t} \, , \label{PSI}
\end{eqnarray}
where we explicitly assume that $a_0 \neq 0$. For our purpose the distribution of participating energies in the initial state should be continuously connected to zero and sufficiently broad, as we will discuss below in more detail. In contrast to \cite{muga}, we are explicitly interested in the survival probability $P(t)$ that the particle remains confined within the well, i.e., 
\begin{eqnarray}
                                P(t) \equiv \int_{-L}^{+L} dx |\Psi(x,t)|^2\,. \label{eq:pp-sur}
\end{eqnarray}
By substituting equation $(\ref{PSI})$ we obtain:
\begin{eqnarray}
                             &&   P(t)= |a_0|^2 I_1+ 2\,a_0\,C_2(t)+C_3(t) \,,  \label{Ptot}   
\end{eqnarray}
where
\begin{eqnarray}
\label{eq:DEF}
                && C_2(t)=\int_0^\infty d E\,a_E\,I_2(E)\,\cos(E\,t)\\
                && C_3(t)=\int_0^\infty d E \int_0^\infty d E'\,a_E\,a_{E'}\,I_3(E,E')\,
                                 \cos[(E-E')\,t] \nonumber \\
       &&  I_1=\int_{-L}^{+L} dx\,|\psi_0(x)|^2, \quad 
              I_2(E)=\int_{-L}^{+L} dx\,\psi_0(x)\,\psi_{E}(x),  \nonumber \\
       &&  I_3(E,E')=\int_{-L}^{+L} dx\,\psi_E(x)\,\psi_{E'}(x). \nonumber
\end{eqnarray}
For the 1D potential of equation~(\ref{eq:1}), the integrals $I_1$, $I_2$ and $I_3$ can be solved analytically,
while this is in general not the case for the 
time-dependent integrals $C_2(t)$ and $C_3(t)$. However, we are interested only in the long-time behavior of $P(t)$, and 
can therefore take advantage of tools from asymptotic theory of Laplace and Fourier transforms \cite{olver,Lighthill}, to obtain the asymptotic form of $P(t)$. 
The functions $C_2(t)$ and $C_3(t)$ can be written as Fourier integrals of the type 
\begin{equation}
C_j(t)= \int_0^\infty dE f_j(E) e^{- i E t} \,,
\end{equation}
with $j=2,3$. By considering the explicit form of the eigenfunctions 
$\psi_0$, $\psi_E$ (see below), one gets for small values of $E$:
\begin{equation}
                I_2 \approx E^{\beta/2}   ~~{\rm and} ~~ I_3 \approx E^{\beta/2} \, {E'}^{\beta/2}\, , \label{smallE}
\end{equation}
with $\beta = (\alpha -3)/2$. Hereafter, we will only consider initial conditions $\Psi_0(x)$ for which $a_0 \neq 0$ and the spectral decomposition 
involves continuum energies that extend down to $E=0$. 

The explicit values of $C_2$ and $C_3$ clearly depend on the initial condition. Hereafter we will give some results for a wide range of initial conditions. For initial states that take non-vanishing values only in the well $[-L,L]$, $a_E \approx E^\beta$ in the limit 
$E \to 0$ \cite{lmm02}. Equation~(\ref{smallE}) furthermore allows to infer 
\begin{equation}
f_2(E) \approx K(\alpha)\,E^\beta ~~{\rm and} ~~f_3(E) \approx H(\alpha)\,E^{\alpha-2}\,,
\end{equation}
for small values of $E$, where $K(\alpha)$ and $H(\alpha)$ are prefactors that depend also on the initial condition. 
If the initial state $\Psi_0(x)$ has nonvanishing values outside the well region $[-L,L]$, the above results still hold,
provided that, for large $|x|\gg L$, $\Psi_0(x)$ exhibits power-law  $\sim  |x|^{-a}$, Gaussian $\sim  \exp(- a x^2)$, or stretched exponential $\sim  \exp(- a x^b)$ decay. The leading contribution to the asymptotic 
long-time behavior of the survival probability is then found to be associated with the $C_2(t)$ term above, and reads:
\begin{eqnarray}
                            &&    P(t) \approx |a_0|^2 I_1+ 2\,a_0\,P_2 {1 \over t^{(\alpha-1)/2}}\, , \quad {\rm{as~}} t \to \infty \, , \label{Pfinal} \\
                            &&    P_2 \approx {1 \over (2\,\pi)^{\beta +1}} 2\,\cos\left( {\pi\over 2} (\beta-1)\right)  \beta ! K(\alpha)\, . \label{P2}
\end{eqnarray}
Through the dependence of $K(\alpha)$ on the spectral expansion of $\Psi_0(x)$, via $f_2$ or $C_2$, the asymptotic decay 
of $P(t)$ also depends on the initial condition. Thus, as anticipated in the introduction, we have shown that the potential of equation $(\ref{eq:1})$ 
induces an algebraic decay in the survival probability $P(t)$, for generic, confined initial states. In contrast,
for an initial state $\Psi_0(x)$ with $a_E=0$ for $E<E_c$ (for some $E_c > 0$), one can show that the survival probability has an exponential cut-off $e^{-E_c t}$ at large $t$, which is a manifestation of the power-law decay arising from continuous spectral components arbitrarily close to the ground state energy.

\begin{figure}
\begin{center}
              \includegraphics[width=0.8\linewidth] {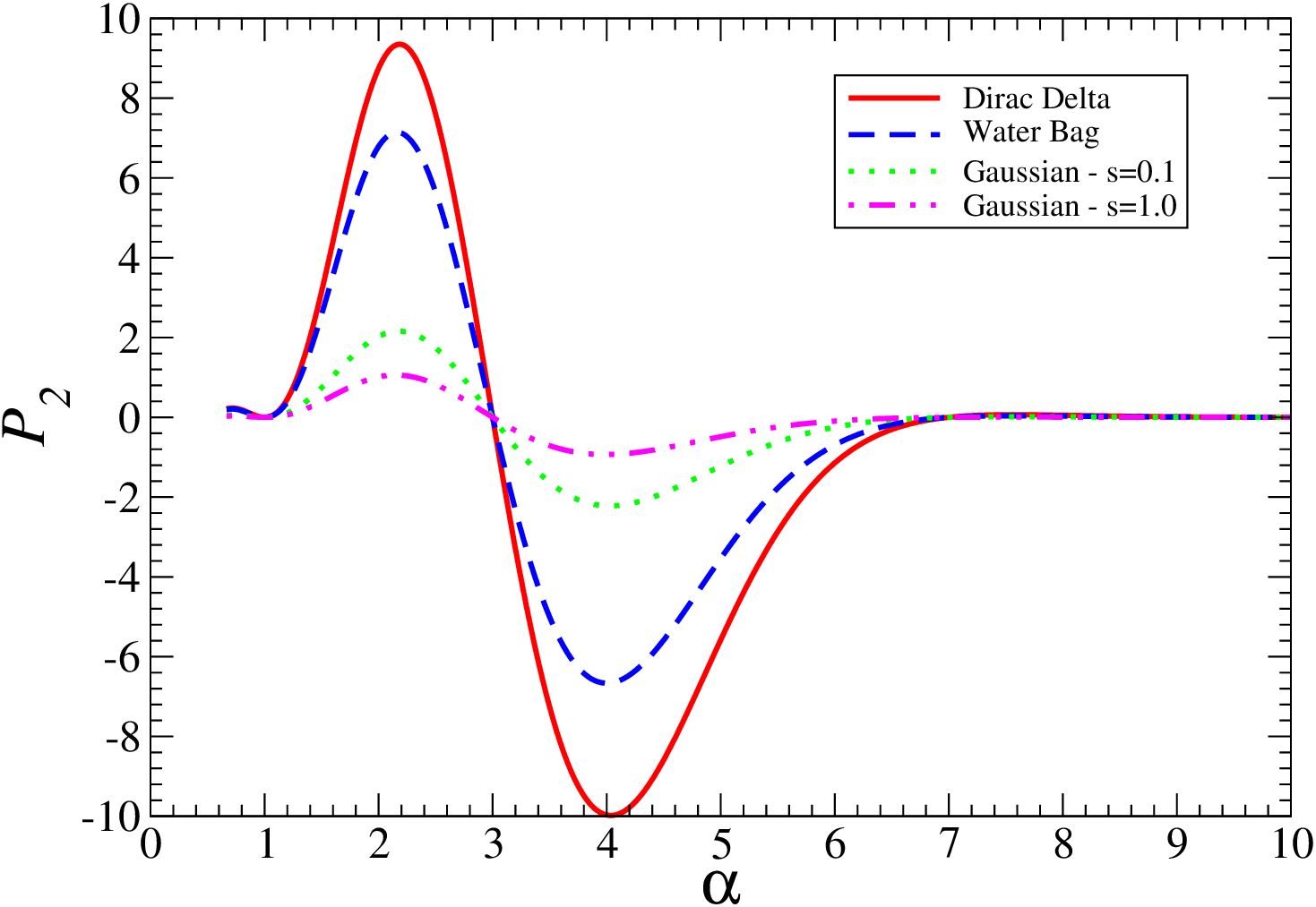} 
	              \caption{Prefactor $P_2$ of the algebraically decaying term in (7), for the potential (\ref{eq:1}) and different initial conditions: (i) Dirac delta $\Psi_0(x)=\delta(x)$ (solid red line); 
	              (ii) water bag  $\Psi_0(x)=1/(2 L)$ in [-L,L] and $\Psi_0(x)=0$ outside (dashed blue line); (iii) Gaussian distribution $\Psi_0(x)=\exp({-x^2/(2\,s)})/\sqrt{2 \pi s}$, with widths $s=0.1$ (dotted green line) and 
	              (iv) $s=1.0$ (dash-dotted magenta line).
	               }   \label{fig:1}
\end{center}   
\end{figure}

Figure~\ref{fig:1} shows $P_2$ for the initial condition $\Psi_0(x)=\delta(x)$. For some values of $\alpha$, e.g. in $\alpha = 1$ and for $\alpha =3$, 
$P_2$ vanishes. The oscillatory behavior of $P_2$ stems from the trigonometric functions 
in $(\ref{P2})$, while the fact that $P_2$ becomes negligible for $\alpha \geq 7$ is induced by 
$K(\alpha)$, and is therefore related to the initial conditions as used in figure~\ref{fig:1}. The oscillatory behavior of $P_2$ may result in a non-monotonic decay of the survival probability (an example will be shown in figure~\ref{fig:3}). In fact, a negative value of $P_2$ implies that the asymptotic value $|a_0|^2 I_1$ is approached from below and therefore the probability derivative must change sign. This is not peculiar of the system we are studying here. In fact, a similar behavior is also observed starting from an usual square well quantum potential $V_S(x)=-V_0$ when $|x| \le L$ and $V_S(x)=0$ when $|x|>L$, with $\Psi_0(x)=\delta(x)$.
For vanishing $P_2$, one has to consider the next leading time dependent term in the asymptotic expansion of $C_2$ and $C_3$. 

In reference \cite{muga}, according to equation (35) therein, a specific initial state was chosen whose expansion coefficients $a_E$ show a power-law behavior, although with an exponent different from ours, for $E \to 0$. This implies a power-law for the survival probability which is different from our result of equation (\ref{Pfinal}). The main difference between our potential of equation~(\ref{eq:1a}) and the 1-D potential considered in \cite{muga} is that the latter is defined only on the positive real axis, while our potential extends over the whole real axis. Consequently, in reference \cite{muga} there does not exist a bound state at $E=0$, leading to a decay without a lower bound, or in other words without control. In contrast, our equation (\ref{Pfinal}) explicitly takes into account $a_0 \neq 0$. Another difference with respect to reference \cite{muga} is that we explicitly study the survival probability inside the quantum well, as defined in equation~(\ref{eq:pp-sur}) and in view of our proposed experiment in section \ref{sect:proposal}, and not the fidelity function considered in \cite{muga}.

\subsection{The role of the potential tails} \label{mupot}

In this section we will show that the mere absence of a gap between the null eigenvalue and the continuum part of the spectrum, is a necessary 
but not sufficient condition in order to observe a power-law decaying survival probability. 

In general, any Schr\"odinger potential that asymptotically decays to zero like $1/x^\mu$ would have a continuum part of the spectrum attached to the null eigenvalue.  Let us now therefore consider the more general case of a Schr\"odinger potential  
\begin{eqnarray}
              V_{\mu}(x)=\left \{\begin{array}{ccc}
                                   -V_0~    &~{\rm{if}}~&~|x| \leq L, \\
                                            &           &                  \\
                                   V_1/|x|^\mu ~&~{\rm{if}}~&~|x| >   L,
                   \end{array} \right.  \qquad \mu<2 \label{VSmu}
\end{eqnarray} 
where $L$, $V_0$, $V_1$ and  $\mu$ are real positive constants. Let us call $\{ \psi_0^{(\mu)}, \psi_E^{(\mu)} \}$ the eigenfunctions associated to such potential. In this case, we can only obtain the eigenfunction associated to the null eigenvalue:
\begin{eqnarray}
               && \psi_0^{(\mu)}(x)= A_0 \sqrt{x} \, K_{\delta}\Bigl(\delta \sqrt{V_1}\,x^{1-\mu/2}\Bigr) \qquad  |x|>L \nonumber \\
               && \delta={1 \over{2 - \mu}}   \,. \label{psi0mu}
\end{eqnarray}
The eigenfunctions relative to the continuum part of the spectrum are not known. The parameters $L$, $V_0$, $V_1$ and $\mu$ can be chosen in such a way that the spectrum contains one single discrete eigenvalue $E_0=0$ and a continuous part for $E>0$. As a result, the parameters $L$, $V_0$, $V_1$ and $\mu$ are not independent. In fact, the continuity of $\partial_x \psi_0$ in $x=L$ provides a relation between them
\begin{eqnarray}
                             && L^{1+\mu/2}\,K_\delta\Bigl(\delta \sqrt{V_1}\,L^{1-\mu/2}\Bigr)\,\Bigl(1+2 L \sqrt{V_0} \tan(L \sqrt{V_0})\Bigr) + \nonumber \\
                             && \hspace{2 truecm} - L^{3/2}\,\sqrt{V_1}\,K_{\delta+1} \Bigl(\delta \sqrt{V_1}\,L^{1-\mu/2}\Bigr)           \\
                             && \hspace{2 truecm} - L^{3/2}\,\sqrt{V_1}\,K_{\delta-1} \Bigl(\delta \sqrt{V_1}\,L^{1-\mu/2}\Bigr)   = 0 \,. \nonumber
\end{eqnarray}
In the following we consider $L$, $V_0$ and $\mu$ as independent parameters and will obtain $V_1$ by numerically solving the above equation. 

We will prove below that the survival probability of the above process of equation $(\ref{VSmu})$ is not in general power-law decaying. In fact, we will show that the $C_2$ term may eventually decay like a power-law for large time values only for very specific initial conditions. Let us make two preliminary observations. First, it is worth mentioning that by using the orthogonality relation:
\begin{eqnarray}
                              \psi_0^{(\mu)}(x) \, \psi_0^{(\mu)}(x')+\int_0^\infty d E \, \psi_E^{(\mu)}(x) \, \psi_E^{(\mu)}(x') = \delta(x-x')
\end{eqnarray}
one can prove the following identity:
\begin{eqnarray}
                                \int_0^\infty d E \, I_2(E) \psi_E(x)= \psi_0^{(\mu)}(x) , \label{cEback} 
\end{eqnarray}
For $I_2$ as defined in equation~(\ref{eq:DEF}).
Secondly, we notice that in order to have $C_2(\tau) \approx 1/\tau^{a+1}$ one must have that $a_E\,I_2(E) \approx E^a$ for small energy values.

Let us now consider the large $x$ behavior of $\int_0^\infty d E I_2(E) \psi_E(x)$. For large values of $x$ the potential vanishes thus $\psi_E^{(\mu)}(x) \approx E^{-1/4} e^{i \sqrt{E} x}$. Therefore, by using equation $(\ref{cEback})$ and equation $(12.01)$ in chapter 3 of reference \cite{olver}, the {\em{ansatz}} $I_2(E) \approx E^a$ would give:
\begin{eqnarray}
                               \int_0^\infty d E \, I_2(E) \psi_E(x) \approx { 1 \over{x^{2 a + 3/2}}}
\end{eqnarray}
This result implies that $\psi_0^{(\mu)}(x)$ should decay according to a power-law, given equation $(\ref{cEback})$. However, by using equation $(\ref{psi0mu})$ one has that:
\begin{eqnarray}
                               \psi_0^{(\mu)}(x) \approx x^{\mu/4} \exp\bigl( - {2 \sqrt{V_1} \over{2 - \mu}}\, x^{1 - \mu/2}\bigl)
\end{eqnarray}
showing that $I_2(E)$ cannot behave as a power-law for small values of $E$ when $\mu<2$. It is worth mentioning that when $\mu=2$, then the above {\em{ansatz}} holds true with $a=(\alpha-5)/4$. In this case, one would get $I(x) \approx x^{1-\alpha/2}$, that is in agreement with the fact that $\psi_0(x) =x^{-\alpha/2}$ for large $x$ values.

We have therefore shown that $I_2(E)$ is not growing like a power-law for small energy values. Therefore in order to have a power-law decay in $C_2(t)$ one has to engineer appropriate initial conditions such that $a_E\, I_2(E)$ behaves like a power-law for small energy values. In conclusion, the absence of an upper bound for the time-scale is a necessary but not sufficient condition in order to observe a power-law decaying auto-correlation function. This also implies that the decay of the survival probability is faster for $\mu < 2$ than for the case with $\mu=2$.

\begin{figure}[t] 
\begin{center}
              \includegraphics[width=0.7\linewidth] {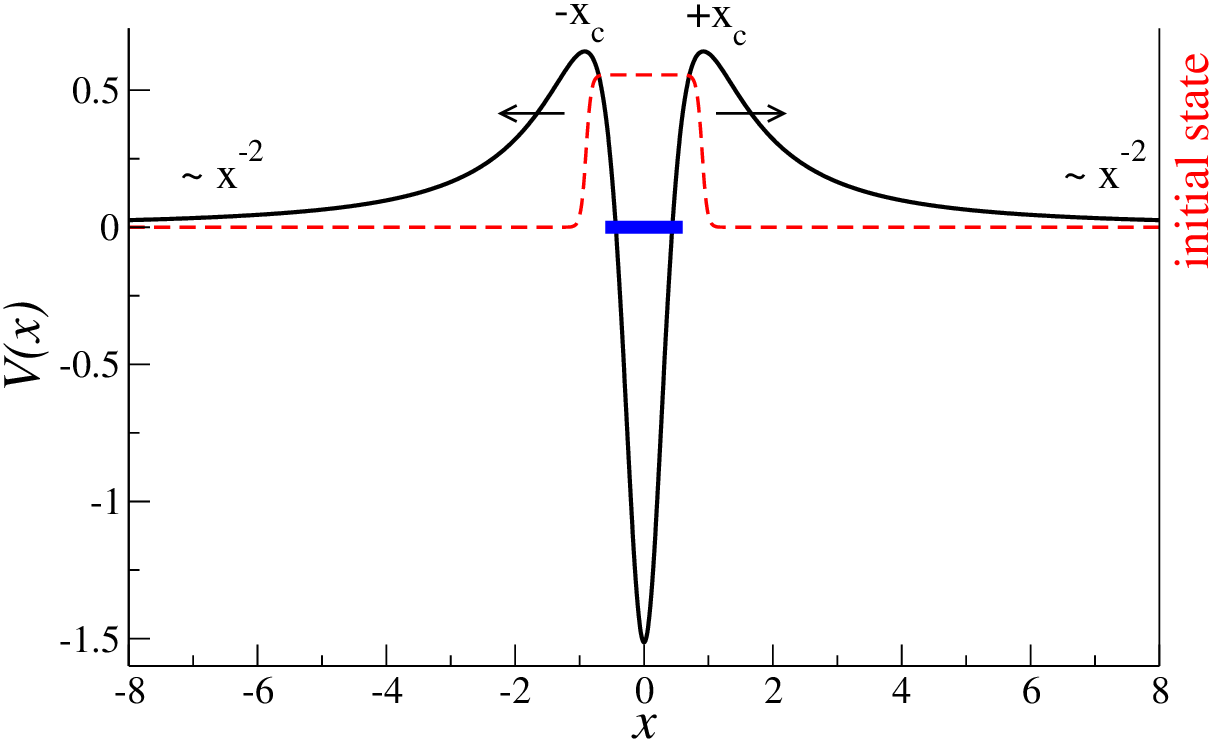}
              \caption{Sketch of the smooth potential $V_{\rm sm}(x)$ for $\alpha = 2.8, \gamma=0.68$ (black solid line), and of the initial state $\Psi_0(x)$ (red dashed line), equation $(\ref{psi_init})$ with $\sigma=30$. The thick blue line shows the ground state energy level within the well.
             }  \label{fig:2}
\end{center}   
\end{figure}

\section{Numerical results} \label{numeric}

\subsection{Numerical Simulations of the model}  \label{num-model}

Because of its discontinuities, it is difficult to reproduce $V_S(x)$ experimentally 
or in numerical simulations. We therefore consider the smoothed version $V_{\rm sm}(x)$ defined 
as  \cite{lm00} 
\begin{eqnarray}
                              && V_{\rm sm}(x) = \frac{\alpha}{4} \frac{x^2(1+\alpha/2)-\gamma^2}{(x^2+\gamma^2)^2}\; , \label{eq:2}  
\end{eqnarray}
in our subsequent numerical tests.  
Also this potential has a continuous spectrum attached to the null eigenvalue, and decays as $x^{-2}$ at $|x|\to \infty$. 
We are interested in the relaxation properties of wave-packets prepared inside the potential well (see figure~\ref{fig:2}). We define the 
survival probability over the region bounded by the potential maxima at $x_c$: 
\begin{eqnarray}
P(t) = \int_{-x_c}^{x_c} dx |\Psi(x,t)|^2 \,, ~~{\rm with}~~~ x_c=\gamma\,((\alpha+6)/(\alpha+2))^{1/2} \,.
\label{eq:psur-def}  
\end{eqnarray}
Our observable is the approach of the survival probability to its asymptotic constant value $P_{\infty}=|a_0|^2 I_1$ described by equation~(\ref{Pfinal}), i.e., the quantity 
\begin{eqnarray}
P_S(t) = |P(t)-P_{\infty}|\,. 
\label{eq:p-sur}  
\end{eqnarray}
Even if we simulate the evolution of the initial wave packet only in one dimension, the numerical computations take quite a long time for several reasons. First of all, we need to propagate sufficiently far out into the tails of the potential, while -- at the same time -- the expansion of the decaying parts is rather fast due to the high energy components of the initial state. This is also the reason why simple absorption methods at the numerical boundary do not work very well since those are typically adapted to absorb just a small window of energies with sufficient precision. Moreover, in order to estimate the asymptotic probability $P_{\infty}$, which is not analytically known for the smooth potential $V_{\rm sm}$, we must propagate considerably longer than shown in the following figures, for which we approximate $P_{\infty}$ by $P(t_{\rm max})$, with $t_{\rm max}= 140 \ldots 200$.

\begin{figure}[t] 
\begin{center}
              \includegraphics[width=0.85\linewidth] {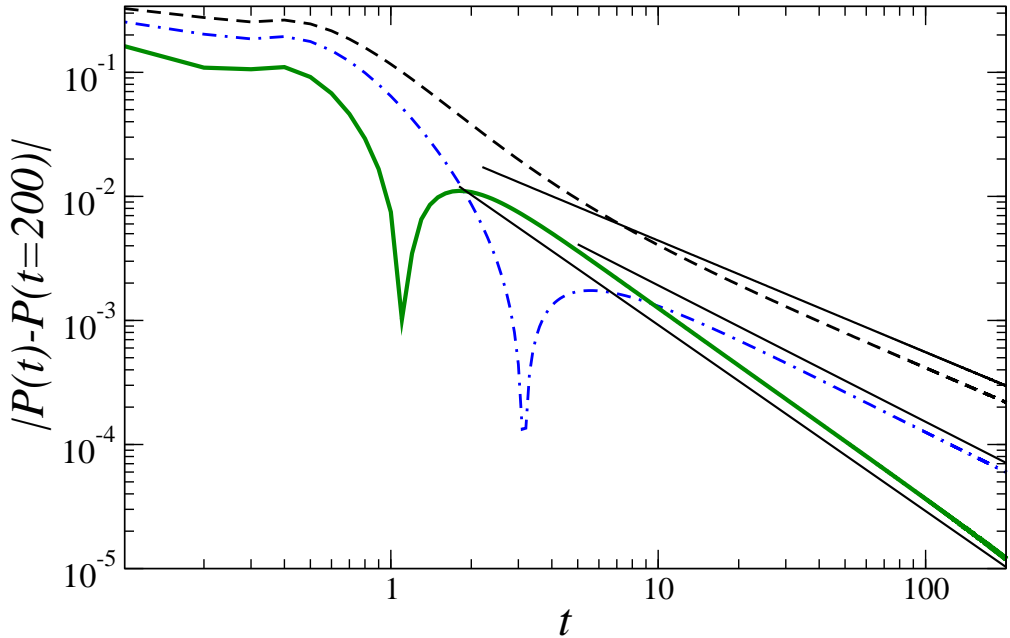} 
              \caption{Survival probability $P_S(t)$,
              for a water-bag like initial state of equation $(\ref{psi_init})$  with $\sigma = 100$ and  $\gamma=0.68$, and increasing values of $\alpha=2.8$ (black dashed line), 
              3.2 (blue dot-dashed), 4 (thick green solid line). The straight black line fits are chosen according to the prediction 
              with exponent $(\alpha - 1)/2$ from equation~(7).
              }                    \label{fig:3}
\end{center}   
\end{figure}

For the actual observation of the predicted asymptotic behavior the precise form of the initial state prepared in the well is not crucial, 
provided it is given by a coherent superposition of energy eigenstates from a continuous energy range including the ground state energy.
We consider a smooth initial state 
\begin{eqnarray}
                              \Psi_0(x) & = \frac{N}{ \left[ 1 + \exp \left( (x-x_c)\sigma \right)\right] \left[ 1 + \exp \left( -(x+x_c)\sigma \right) \right] }  \, , \label{psi_init}
\end{eqnarray}
where $N$ is a normalization constant. The problem defined by the potential of equation~(\ref{eq:2}) is not exactly solvable unfortunately. Therefore the expansion coefficients  $\{a_0, a_E\}$ are not known. This prevents us from giving an analytical solution for the survival probability, given the above initial state. We are therefore forced to perform numerical simulations, as detailed below. However, the fact that the expansion coefficients are not known makes it also difficult to estimate what is the initial population of the states in the continuum part of the spectrum. We can, however, estimated the mean energy $\bar{E}$ of the initial state of equation (\ref{psi_init}). By using the parameters of figure~\ref{fig:2}, we get for instance $\bar{E} \approx 2.6$.

The initial state (\ref{psi_init}), shown in figure~\ref{fig:2}, is numerically propagated in real time using an implicit norm-preserving Crank-Nicolson integration scheme \cite{Pisa}, which also controls the boundary conditions very well \cite{paul}. Figure~\ref{fig:3} reports the results of the numerical simulations. One can observe a clear algebraic decay as predicted by equation~(7), for various values of $\alpha$ which induce different algebraic decay exponents. The power-law decay emerges the earlier, the larger the weight factor $P_2$ in (7). The asymptotic value $P_\infty$ is approached from above ($\alpha = 2.8$) or from below ($\alpha = 3.2, 4$), depending on this factor's sign (compare figure~1). In the latter case, a change of sign of $P(t)-P_\infty$ at finite times induces the discontinuities in the first derivative on the logarithmic scale of figure~\ref{fig:3}. For fixed $\alpha$, the power-law scaling regime may be further enhanced by adapting the precise form of the initial state (e.g. by the above parameter $\sigma$), and also by changing the second parameter $\gamma$ of the potential of equation~(\ref{eq:2}). Both enter our approximate formula for the prefactor $P_2$, see equation~(\ref{P2}) and figure~\ref{fig:1}. 

The figure explicitly shows how the decay of the survival probability may not necessarily be monotonic as, for example, in the green and blue curves. The effect is magnified by the logarithmic scale. As mentioned in section \ref{theory-A}, this is connected with the pre-factor $P_2$, which inverts its sign, as shown in figure~\ref{fig:1}. We do not have a precise physical explanation of this oscillatory behavior. Indeed, the time evolution of $\psi_E$ is highly nontrivial. This induces an alternation between (i) phases when the the bulk of the state is outside the well, and therefore we observe a depletion of the survival probability, and (ii) phases when there is a re-entrance of the state within the well, which causes a partial restoration of the survival probability. In fact, no matter its mean energy, $\Psi_0$ is the linear superposition of eigenstates some of which have very low energies, in particular smaller than $V_S(x_c)$. Exactly these low energy components give rise to the observed power-law in the survival probability after the transient, as we have shown in section \ref{theory-A}. The oscillatory transient behavior is also present for a simple square-well potential, which tells us that the transient should mainly be caused by the high-energy components contributing to the dynamics while the initial state relaxes in the well.

\subsection{The role of the tail in the potential} 

In the following we will show that the predicted power-law decay is robust with respect to realistic experimental situations. The only parameter which needs to be controlled very well is the exponent of the potential tails, i.e. $\mu$ in the asymptotics of the potential scaling as $\sim x^{-\mu}$. We use the form
\begin{eqnarray}
                              && V_{\rm sm, \mu}(x) = \frac{\alpha}{4} \frac{x^2(1+\alpha/2)-\gamma^2}{(x^2+\gamma^2)^{\mu/2+1}}\; , \label{eq:2-mu}  
\end{eqnarray}
which for $\mu=2$ recovers $V_{\rm sm}(x)$ from equation~(\ref{eq:2}).

\begin{figure}[t] 
\begin{center}
              \includegraphics[width=0.8\linewidth] {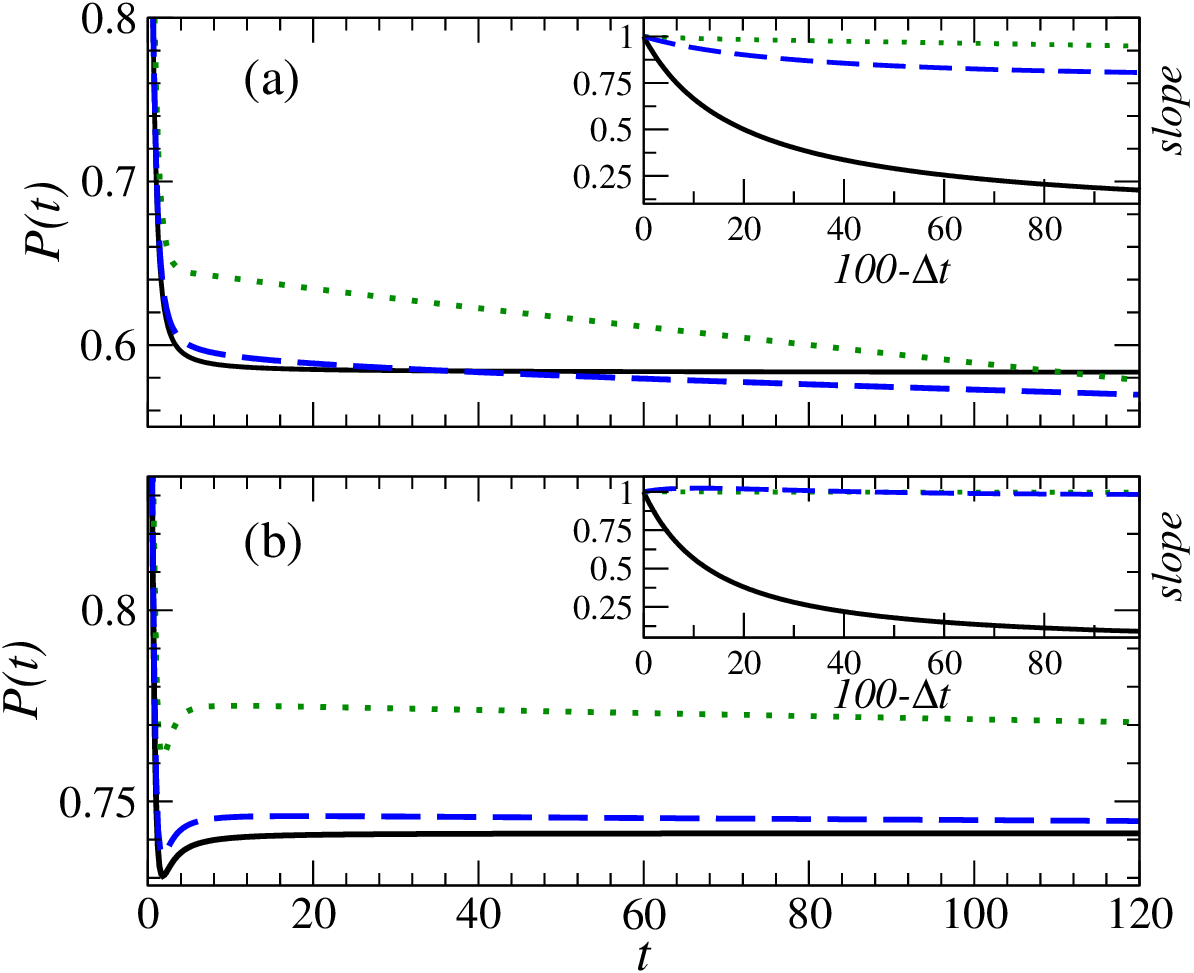}
              \caption{Survival probability $P(t)$, for a water-bag like initial state of 
              equation $(\ref{psi_init})$  with $\sigma = 60$, for well parameters $\gamma=0.68, \alpha=2.8$ (a) and $\alpha=4$ (b), 
              and asymptotics with exponents $\mu=1.5$ (green dotted line), $\mu=1.9$ (blue dashed line) and $\mu=2$ (black solid line), respectively.
              The insets show, for $t=20\ldots120$, the decay rates or slopes defined in (\ref{eq:p-sur-slope}): $(P(120)-P(t))/\Delta t$ 
              vs. $100 - \Delta t = t - 20$, with $\Delta t \equiv 120 - t$.
              Hence, we observe a relaxation with a fast decreasing slop as $\Delta t \to 0$, in the special case of
              $\mu=2$. In all the other cases the slope remains almost constant over the shown times, implying a much faster decay with time.
              }                    \label{fig:4new}
\end{center}   
\end{figure}

For $\mu \neq 2$, the dynamics of the system will be qualitatively very different, as we show in figure~\ref{fig:4new} directly for the survival probability. Instead of an asymptotically slow saturation toward the ground state within the potential, we observe a fast decay -- with approximately constant slopes -- for all values $\mu < 2$ along the same time scales as in figure~\ref{fig:3}. 

As described above around equation~(\ref{eq:p-sur}), it is computationally hard to estimate the saturation value of the survival probabilities, since the ground states are not known analytically and the data for $\mu \neq 2$ implies that a reliable numerical estimate will be possible only after much longer propagations than shown in the figures. This may cause the wrong impression suggested by the main panels of figure~\ref{fig:4new}: the ground state is approached in shorter {\em absolute} time for $\mu=2$ as compared to the other values of $\mu$, yet what counts is the {\em rate} of approach which is always smaller for $\mu = 2$. This is what we intend as {\em 'slower'} decay in the latter case and the reason for plotting the slopes defined by
\begin{eqnarray}
\frac{\left(P(120)-P(t)\right)/ \Delta t}{\left(P(120)-P(20)\right)/100} \,,
\label{eq:p-sur-slope}  
\end{eqnarray}
in the insets, with $\Delta t \equiv 120 - t$. In order to obtained comparable values, we divided by the extremal values, see the denominator of (\ref{eq:p-sur-slope}), which all occur at the minimal time $t=20$ (chosen after non-universal transients at still smaller times). Only in the special case of $\mu=2$, the relaxation to the steady-state value becomes much slower with time, see the fast decreasing slopes in the insets (black solid line). The shown numerical results confirm the theoretical discussion above in section \ref{mupot}, predicting in essence a faster approach to the ground or equilibrium state for $\mu < 2$. Consequently, a possible experiment should control the exponent to be equal to two at least at 2-3 significant digits, in order to observe the power-law decay discussed in sections \ref{theory-A} and \ref{num-model}.

\subsection{Experimental realization of the model} 
\label{sect:proposal}

Let us conclude with an experimental protocol to test our prediction (\ref{Pfinal}). We have in mind a Bose-Einstein condensate prepared in an optical trap which then is exposed to a potential of the form of equation~(\ref{eq:2}) while the trap is relaxed. The potential may hereby be created optically, for instance, with a fast moving laser beam \cite{davidson} or by holographic techniques \cite{holo}.

\begin{figure}[t]
\begin{center}
              \includegraphics[width=0.85\linewidth] {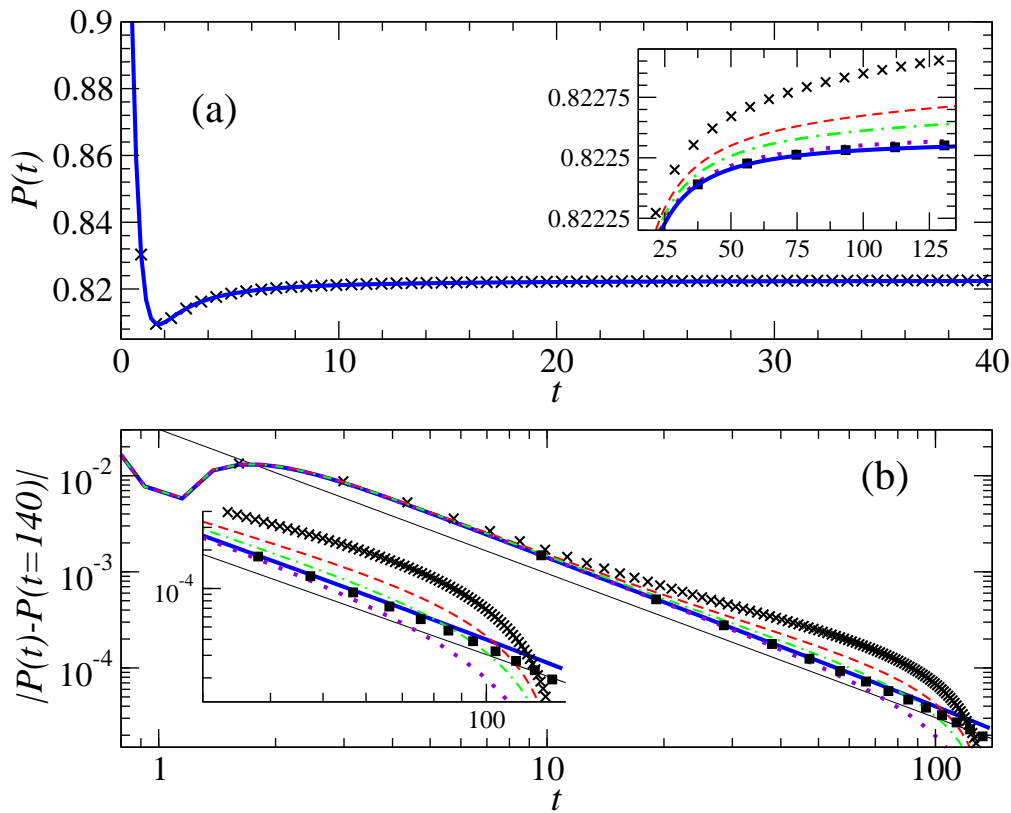}
              \caption{
               (a) Survival probability for a confining potential with $\alpha=4$ (i.e. $x_c  = 0.88 $), and otherwise the same parameters as in figure~\ref{fig:3}. The particle is here prepared in a Gaussian initial state 
               with a spatial 
               width $\sigma_{\rm trap,\,i}=0.5$, and evolves in the potential (\ref{eq:2}) without further perturbation (thick blue lines), or in the presence of an additional harmonic trap of width 
               	$\sigma_{\rm trap,\,f}=100$ (crosses), 150 (red dashed), 200 (green dot-dashed), 400 (dotted), and 800 (black squares). Inset: long-time behavior. (b) Data from (a) 
                shown in the form of $P_S(t)$, equation (\ref{eq:p-sur}), for direct comparison with 
		figure~\ref{fig:3} and the power law of equation~(7), with $(\alpha - 1)/2=1.5$ (thin solid line).}
              \label{fig:5}
\end{center}   
\end{figure}

A Gaussian initial state of the form 
\begin{equation}
\Psi_0(x) = N \, \exp(-x^2/(4\sigma_{\rm trap,\,i}^2))
\end{equation}
is initially prepared in a harmonic trap with characteristic oscillator length $\sigma_{\rm trap,\,i}$. $N$ is a normalization constant. Then the smooth potential (\ref{eq:2}) is switched on, while we switch off or relax the trap abruptly to a shallow confinement characterized by the harmonic oscillator length $\sigma_{\rm trap,\,f} \gg \sigma_{\rm trap,\,i}$. In the case of figure~\ref{fig:5} we consider values from $\sigma_{\rm trap,\,f}=100$ to $\sigma_{\rm trap,\,f}=800$, and $\sigma_{\rm trap,\,i}=0.5$. In the absence of a trap (thick blue lines), the Gaussian initial state exhibits a behavior similar to that of the smooth water-bag state from above, equation~(\ref{psi_init}). A shallow trap with large $\sigma_{\rm trap,\,f}$ manifests at long times, by an exponential cut-off of $P_S(t)$, as shown in figure~\ref{fig:5}, since the potential's asymptotics are changed by the trap. This induces a spectral gap between the ground state energy and the continuum component, while it is specifically the absence of the gap which is responsible for the algebraic decay, as discussed in section \ref{theory-A}. The steeper the confining trap potential, the larger the trap-induced spectral gap, and the shorter the time interval over which an algebraic decay can be observed, before the exponential cut-off. Indeed, we observe such a continuous degradation of the asymptotic law with exponent $(\alpha - 1)/2=1.5$ when making the additional confinement steeper. However, the power-law trend is still clearly visible, over at least one order of magnitude, even in the presence of the steepest trap (with $\sigma_{\rm trap,\,f}=100$ in figure \ref{fig:5}). 

\begin{figure}[t]
\begin{center}
              \includegraphics[width=0.85\linewidth] {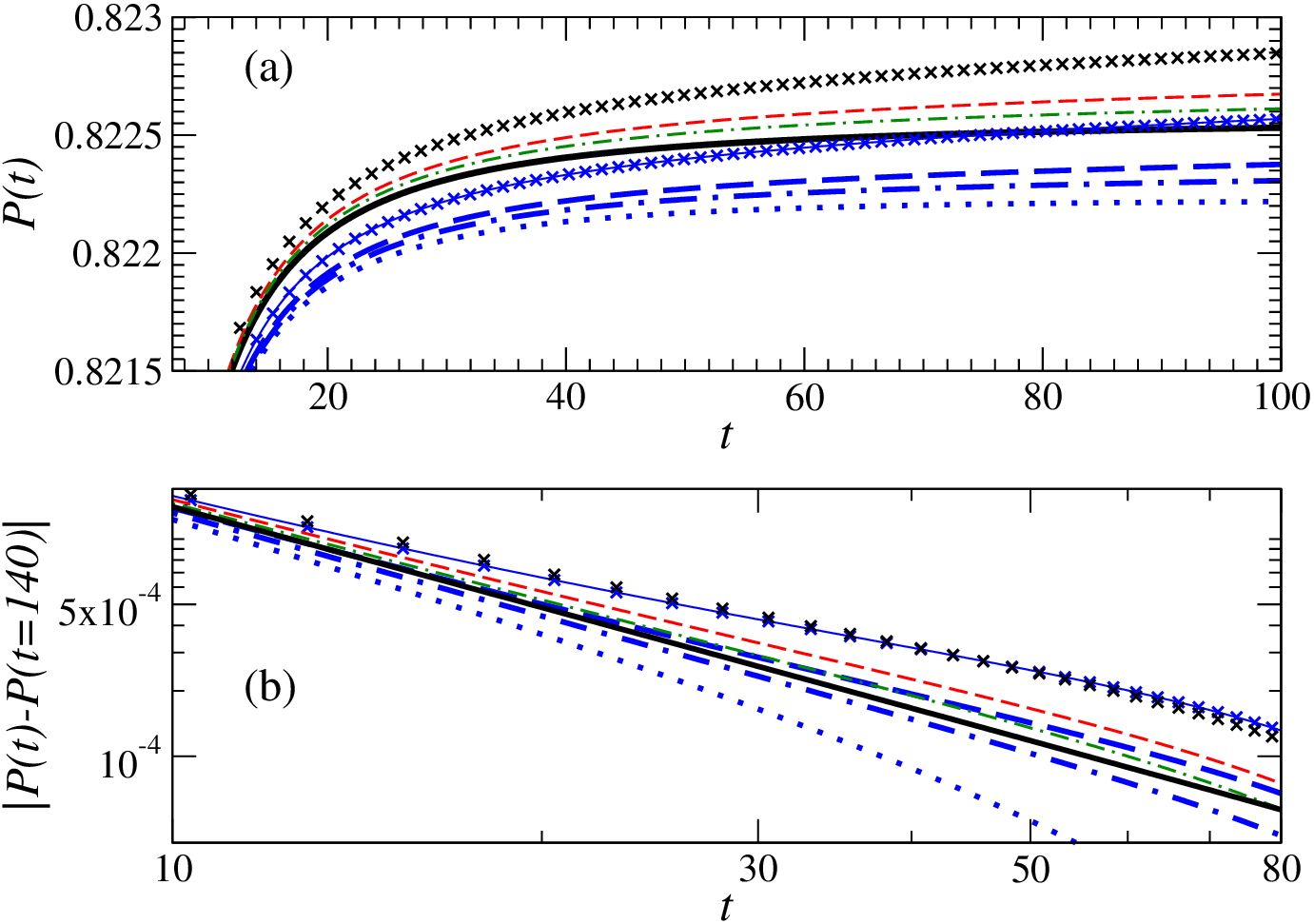}
              \caption{
                   Direct and shifted survival probabilities $P(t)$ (a) and $P_S(t)$ (b), respectively,
                   taken from figure~\ref{fig:5} without any trap (thick black solid line) and for $\sigma_{\rm trap,\,f}=100$ (crosses), 
                   150 (thin red dashed), 200 (thin green dot-dashed) for $g=0$, in comparison with the same cases and a weak 
                   repulsive nonlinearity $g=5\times10^{-3}$ and no trap (blue dotted) or 
                   $\sigma_{\rm trap,\,f}=100$ (crosses connected by thin line), 150 (thick blue dashed), 200 (thick blue dot-dashed). 
                   }
              \label{fig:6}
\end{center}   
\end{figure}

One may ask about the role of interactions between the weakly interacting atoms of a Bose condensate. Those effectively scale in a mean-field approximation with the number of atoms, which may be controlled and possibly be reduced \cite{bdz}. We modelled the evolution of an initially well confined condensate, following the above protocol, using a one-dimensional Gross-Pitaevskii equation. In this approach the interactions are taken into account by a nonlinear density-dependent term in the Hamiltonian \cite{bdz,Pisa,paul}. A weak repulsive nonlinearity may actually stabilize the evolution and -- to some extent -- reverse the effect of a weak confinement during the relaxation (c.f. \cite{naegerle} for a similar effect). This can be seen in our final figure~\ref{fig:6}, where the prefactor of the nonlinear term is denoted by $g$ in our dimensionless units\footnote{Our dimensionless potentials are given in units of $energy/m^2$. The three-dimensional atom-atom interaction strength can be reduced to an effective one-dimensional parameter, provided a strong transverse/radial confinement is experimentally achieved \cite{bdz}. This latter parameter $g_{\rm 1D}=2\hbar\omega_{\rm rad} a_S N_a$, with the scattering length $a_S$, the radial confinement frequency $\omega_{\rm rad}$ and the number of atoms in the Bose condensate $N_a$, can be expressed without dimensions using, e.g., just the single scale given by the radial confinement: $E_0=\hbar\omega_{\rm rad}$ and $x_0=\sqrt{2/\hbar\omega_{\rm rad}M}$, where $M$ is the single atom mass. This gives $g=g_{\rm 1D}/(E_0x_0)=2N_aa_S/x_0$. Here the wave function is normalized to one as in our single particle computations otherwise used in the paper.}.

\section{Conclusions} \label{conclusions}

In summary, we propose a setup for which one may observe a power-law decay of the asymptotic survival probability $P_S(t)$ in a controlled manner. Such an anomalous decay is readily realized by preparing a quantum wavepacket with a sufficiently broad energy range around the ground state energy $E=0$ in the potential wells given by the equations (\ref{eq:1}) and (\ref{eq:2}). The slow algebraic relaxation toward this equilibrium state arises from the spectral properties and the population of eigenmodes of the system. The latter are encoded in the coefficient $C_2$ defined in equation~(\ref{Ptot}). Hence, the predicted behavior is of purely quantum origin (not induced, however, by dynamical or Anderson localization \cite{anderson,casati79} as discussed in \cite{casati99}, nor by semiclassical arguments \cite{grebogi}). The exceptional control of state-of-the-art experiments with ultracold atoms \cite{billy,saw,selim,davidson,toulouse,holo} offers the possibility to observe our predictions, following, e.g., our protocol for obtaining the data of figures~\ref{fig:5} and \ref{fig:6}.

\ack
Support by the Heidelberg Center for Quantum Dynamics, Global Networks Mobility Measures, Frontier Innovation Fund, and the DFG through the HGSFP (Grant No. GSC 129/1) and FOR760 is gratefully acknowledged.

\appendix
\section{Details on the analytical solution}
\label{sol}

Following the references \cite{lmm02,farago}, we review here the analytical solution of the problem given by
the potential of equation (\ref{eq:1a}). The eigenfunctions for the ground state are:
\begin{eqnarray}
                             &&     \psi_0=A_0/x^{\alpha \over{2}} \quad \rm{for}\, |x|>L,\\
                             &&     \psi_0=B_0 \,\cos(\sqrt{V_0}\,x) \quad \rm{for}\, |x|<L,
 \end{eqnarray}
with $\alpha=\sqrt{1 + 8 V_1}-1$. By imposing the continuity in $x=\pm L$ one gets:
\begin{eqnarray}
                                 B_0 =A_0 {{L^{-\alpha/2}}\over{\cos \bigl( L \sqrt{V_0}\bigl)}}.
\end{eqnarray}
By imposing that the ground state is normalized to unity one gets:                              
\begin{eqnarray}
                                 A_0 =L^{(\alpha-1)/2}\,\Bigl( {2 \over {\alpha-1}} +
                                                                                     \sec \bigl( L \sqrt{V_0}\bigl)^2 +
                                                                                     {{\tan \bigl( L \sqrt{V_0} \bigl)} \over { L \sqrt{V_0} }}  
                                                                          \Bigl)^{-1/2}.
\end{eqnarray}
The continuity of the first derivative in $x=\pm L$ is ensured by Eq. $(\ref{eq:1b})$. The odd eigenfunctions for the continuum part of the spectrum are: 
\begin{eqnarray}
                             &&     \psi_E^{(odd)}=a_E\,\sqrt{x} J_\nu(\sqrt{E}\,x)+b_E\,\sqrt{x}\, Y_\nu(\sqrt{E}\,x)     \quad \rm{for}\, |x|>L,  \\
                             &&     \psi_E^{(odd)}=d_E\,\sin(\sqrt{V_0+E}\,x) \quad \rm{for}\, |x|<L,
 \end{eqnarray}
where $\nu=(\alpha+1)/2$. By imposing the continuity in $x=\pm L$ one gets:
\begin{eqnarray}
                                 d_E =\sqrt{L} \, {1 \over \cos\bigl( \sqrt{E+V_0} \bigl)}\,
                                            \Bigl(
                                                      J_{\nu} \bigl( L \sqrt{E} \bigl) \, a_E+ 
                                                      Y_{\nu} \bigl( L \sqrt{E} \bigl) \, b_E
                                            \Bigl).
\end{eqnarray}
By imposing the normalization condition $\int dx \, \psi_E(x) \psi_{E'}(x) = \delta(E-E')$ one gets:
\begin{eqnarray}
                           &&      a_E ={1 \over \sqrt{2}}\, \cos \bigl( \Lambda_\nu + \Lambda(E)\bigl), \quad \quad 
                                      b_E ={1 \over \sqrt{2}}\, \sin \bigl( \Lambda_\nu + \Lambda(E)\bigl) .
\end{eqnarray}
By imposing the continuity of the first derivative in $x=\pm L$ one gets:
\begin{eqnarray}
                           &&      \Lambda(E)   =- \Lambda_\nu+ 
                                                                    \arctan\Bigl( {{F(E) J_\nu(\sqrt{E} \,L)+ \sqrt{E}\, J_{1+\nu}(\sqrt{E} \,L) }
                                                                                             \over
                                                                                            {F(E) Y_\nu(\sqrt{E} \,L)+ \sqrt{E}\, Y_{1+\nu}(\sqrt{E} \,L) }} 
                                                                                 \Big), \\
                           &&  \nonumber \\                                                      
                           &&     F(E)=\cot \bigl( L \, \sqrt{E+V_0} \bigl)\, \sqrt{E+V_0} -{1 \over {2 L}} \,  \bigl( 1+\sqrt{1 + 8 V_1}\bigl).
\end{eqnarray}

\section*{References}

\end{document}